# A metasurface composed of 3-bit coding linear polarization conversion elements and its application to RCS reduction of patch antenna

**Xiangkun Kong**[1,2,*], **Qi Wang**[1], **Shunliu Jiang**[1], **Lingqi Kong**[1], **Yuan Jing**[1], **Xiangxi Yan**[1], **and Xing Zhao**[1,+]

[1] Key Laboratory of Radar Imaging and Microwave Photonics, Nanjing University of Aeronautics and Astronautics, Nanjing, 210016, P.R. China
[2] State Key Laboratory of Millimeter Waves of Southeast University, Nanjing, 210096, P.R. China
*xkkong@nuaa.edu.cn

## ABSTRACT

In this paper, a metasurface composed of 3-bit coding linear polarization conversion elements and its application to RCS reduction of the patch antenna is intensively studied. At first, 3-bit coding metamaterials are constructed by a sequence of eight coded unit cells, which have a similar cross-polarized reflected amplitude response and gradient reflected phase responses covering 0~2π, respectively. Equivalent circuit models (ECMs) of these unit cells are created to describe their electrical behavior for the two linear incident polarizations at the same time. Then, a patch antenna is integrated on the 3-bit metasurface, of which the elements are placed with a 2-dimensional linear coding sequence. The metal square ring is set around the patch antenna to protect it from the disturbance of metasurface. Both the simulation and experiment results demonstrate that the designed metasurface can primarily reduce the antenna RCS at a broadband, while the antenna performances are not degraded significantly.

## Introduction

The antenna is always one of the dominant scatters of a carrier[1]. So it is essential to make antenna in low-observable or radio-transparent states in stealth and anti-stealth technology. Traditional RCS reduction approaches, including loading radar absorbing materials[2], shaping the antenna structure[3], are not suitable for the general antenna due to the sacrifice of the radiation performances. In order to keep a better balance between the antenna performances and RCS reduction, the frequency selective surfaces (FSSs) are applied to reduce the RCS by controlling the energy that reaches the antenna[4]. Unfortunately, it cannot deal with the in-band RCS reduction.

With the development of metasurfaces, also called the two-dimensional version of metamaterials[5], more attention is paid on the in-band RCS reduction of antennas by using metasurfaces to redirect the scattered energy. In[6-8], various artificial magnetic conductors (AMCs) are placed in checkerboard arrangement to reduce the monostatic RCS based on the scattering cancellation. However, this kind of chessboard metasurfaces (CMs) suffers from the narrow band. The chessboard polarization conversion metasurfaces (PCMs) are used to expand the bandwidth of RCS reduction, in[9-11]. Aside from the CMs, phase gradient metasurfaces (PGMs) are also utilized to reduce the RCS of antennas[12-13], based on the generalized Snell-Descartes's law[14]. In[15], a technique, which represents the generalized Snell-Descartes's law of reflection in the form of an array factor, is proposed and converges the two RCS-reduction methods: CMs and PGMs. Both the two methods would cause a sharp RCS increase in some directions, which is harmful for the invisibility of antennas under bistatic detection. Cui *et al.* proposed the concept of coding metamaterials[16], which is very meaningful for realizing excellent scattering reductions because it simplifies the distributions of a set of artificially designed scatters in the form of coding sequences. This method makes a bridge between specific optimization algorithms and the required electromagnetic scattering effect[17-19].

The cornerstone of coding metasurfaces is the design of multi-bits elements. Most multi-bits elements have similar amplitude responses and different phase responses. Moreover, they cannot change the polarization state of the electromagnetic wave. Linear polarization conversion phase gradient metasurfaces are introduced in[5, 20], which could provide polarization states as design freedom to multi-bits elements. To the best of our knowledge, there is a lack of the antenna combined with metasurfaces composed of multi-bits linear polarization conversion elements.

In this paper, a metasurface composed of 3-bit coding linear polarization conversion elements and its application to RCS reduction of the patch antenna is intensively studied. Firstly, 3-bit coding metamaterials are constructed by a linear phase sequence of eight unit cells. And, ECMs of these unit cells are built to describe their electrical behavior. Then, a patch antenna is integrated on the metasurface, of which the elements are placed with 2-dimensional linear coding sequence. At last, the antenna is fabricated and tested. Both the simulated and measured results indicate the proposed antenna has strong RCS reduction compared with the reference antenna over a wide frequency band, while its radiation performances are not obviously degraded.

## 3-bit coding unit cells and equivalent circuit model

The unit cell consists of a top metallic periodic structure, a dielectric layer (F4B, $\varepsilon_r$ =2.65, tan δ=0.001) with a height of h and a bottom metallic ground plane, as shown in Fig 1(a). The top metallic periodic structure is comprised of the long and short metallic arc wires, as well as the metallic disks. The structure has already been proved to achieve wideband polarization conversion[21]. Main resonances are excited by the long arc metallic wire and the metallic disk. Moreover, the change of the short arc wire parameters would produce small perturbations on the electromagnetic behavior of the structure, which is beneficial to get ideal cross-polarized reflected phase.

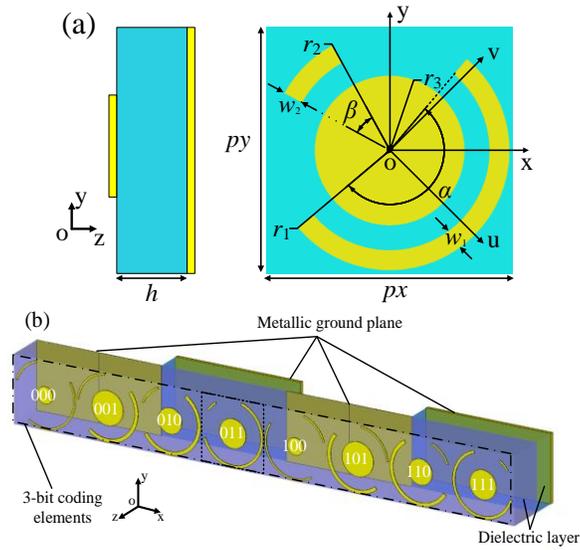

**Figure 1.** (a) The unit cell geometry; (b) 3-bit coding unit cells.

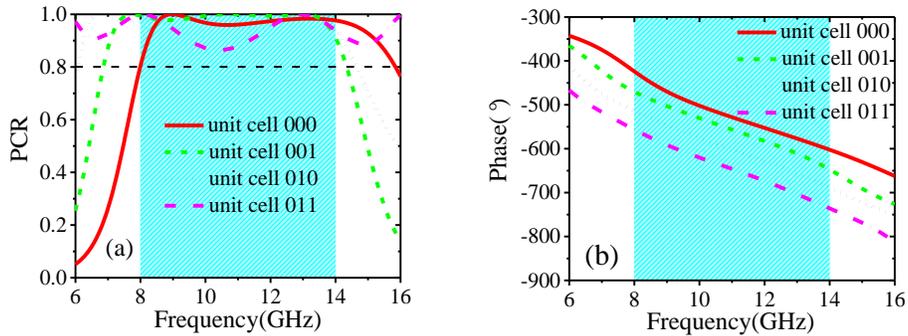

**Figure 2.** Under y-polarized plane wave incidence, (a) simulated PCR of four polarized conversion unit cells; (b) simulated cross-polarized reflection phase of four polarization conversion unit cells.

By changing some parameters, as shown in Table.1, four unit cells are introduced by numerical simulations to form nearly a π phase shift coverage for the cross-polarized reflected light. These unit cells are coded as 000,001,010,011, as shown in Fig.1 (b), which represent 0, -π/4, -π/2, -3π/4 cross-polarized reflected phase response, respectively. The same geometry parameters of these unit cells are as follows: px=py=8 mm, r1=r2=3.9 mm, w2=0.2 mm. Full-wave simulation is performed to investigate the reflection property of these unit cells by using CST microwave studio with periodic boundary conditions in x- and y- directions and open conditions along the z-direction. Polarization conversion ratio (PCR) is defined as:

$$PCR = r_{yx}^2 / (r_{yx}^2 + r_{xx}^2) \ or \ r_{xy}^2 / (r_{xy}^2 + r_{yy}^2) \tag{1}$$

Where $r_{xx}$ or $r_{yy}$ represents the co-polarized reflection coefficient. $r_{yx}$ or $r_{xy}$ represents the cross-polarized reflection coefficient under x- or y-polarized incident wave. From Fig.2 (a), it is found that the PCR of all unit cells is higher than 80% from 8 to 14 GHz under the incidence of the y-polarized wave. Moreover, their cross-polarized reflected phase responses are linear with the coding sequence and the phase difference between two adjacent coding unit cells keeps 45 ° during the same frequency band, as shown in Fig.2 (b). The other four unit cells can be gotten by mirroring previous four unit cells along y-axis and noted as 100,101,110,111, showing -π, -5π/4, -3π/4, -7π/4 cross-polarized reflected phase response and the same amplitude response as 000,001,010,011, respectively.

Table I. Final optimized geometry parameters of four unit cells

| N | α(°) | β(°) | $w_1$(mm) | $r_3$(mm) | h(mm) |
|---|---|---|---|---|---|
| 000 | 155 | 55 | 0.2 | 1.0 | 3.5 |
| 001 | 165 | 80 | 0.4 | 2.0 | 3.5 |
| 010 | 205 | 45 | 0.4 | 1.5 | 4 |
| 011 | 235 | 20 | 0.3 | 1.7 | 4 |

To further understand the polarization conversion mechanism of these unit cells, we set up a four ports network according to the reference[22], as shown in Fig.3 (a). S11 or S44 is defined as the co-polarized reflection coefficient, and S41 or S14 is defined as the cross-polarized reflection coefficient under the normal incidence of x- or y- polarized wave propagating along – z-axis, respectively. Due to the metallic ground sheet, port 2 and port 3 are loaded by conducting short. Z0 is the impendence of free space (Z0=377 Ω). The F4B layer is equivalent to the transmission line with length h and impendence Zr. The top metallic structure can be described by connection quadripole in transmission parameters Tc, which is noted as[22]:

$$\mathbf{T}_c = \begin{pmatrix} A & B \\ C & D \end{pmatrix} = \begin{pmatrix} \frac{Y_u + Y_v}{Y_u - Y_v} & \frac{2}{Y_u - Y_v} \\ \frac{2Y_u Y_v}{Y_u - Y_v} & \frac{Y_u + Y_v}{Y_u - Y_v} \end{pmatrix} \tag{2}$$

Where $Y_u$ or $Y_v$ is the parallel admittance under the normal incidence of u- or v-polarized wave. The equivalent circuit of Tc must exhibit the architecture in Fig. 3(b). According to the Foster representation and from the perspective of simplification, Yu can be seen as a capacitance Cm in parallel with an inductance $L_v$, which is in series with a capacitance Cv. At the same time, $Y_v$ can be expanded in a similar condition in Fig.3(c). Compared with the ECM of cut wire[23], we add an extra capacitance because of the complexity of the designed structure. The ECMs of unit cells are set up and simulated using ADS software.

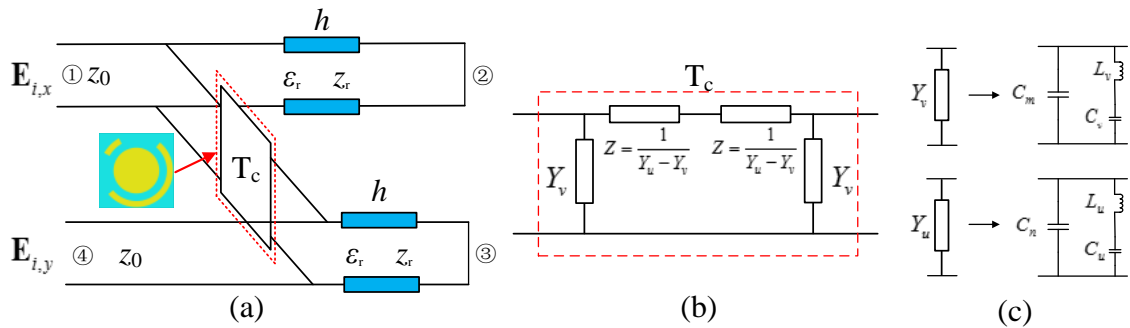

**Figure 3.** (a) The ECM of the unit cell shown in Fig. 1. (b) Pi-representation of the connection quadripole; (c) Equivalent circuit of the unit cell for v-and u-polarized wave.

The final synthetic values of inductance and capacitance are as follows: for 000 element, $C_m$=8.01e-15 F, $C_v$=3.37e-14 F, $L_v$=6.06e-9 H, $C_n$=1.28e-14 F, $C_u$=3.49e-11 F, $L_u$=5.92e-8 H; for 001 element, $C_m$=3.00e-15 F, $C_v$=4.16e-14 F, $L_v$=5.78e-9 H, $C_n$=3.26e-14 F, $C_u$=3.49e-12 F, $L_u$=2.55e-8 H; for 010 element, $C_m$=2.40e-14 F, $C_v$=5.79e-11 F, $L_v$=5.53e-9 H, $C_n$=1.88e-

11 F, $C_u$=4.50e-15 F, $L_u$=5.97e-9 H; for 011 element, $C_m$=1.20e-14 F, $C_v$=6.83e-11 F, $L_v$=7.15e-9 H, $C_n$=3.05e-14 F, $C_u$=1.60e-14 F, $L_u$=7.30e-9 H. The Fig. 4 and Fig.5 reveal that the simulation results of ECMs and CST are in a great agreement.

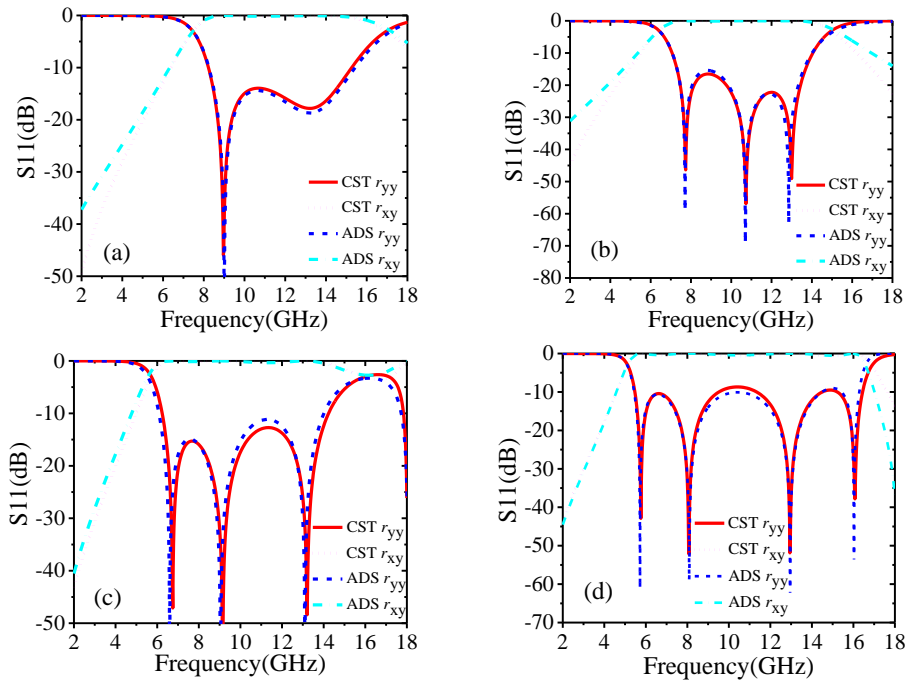

**Figure 4.** Circuital (ADS) and electromagnetic simulated (CST) cross-polarized reflected amplitude responses (a) unit cell 000; (b) unit cell 001; (c) unit cell 010; (d) unit cell 011.

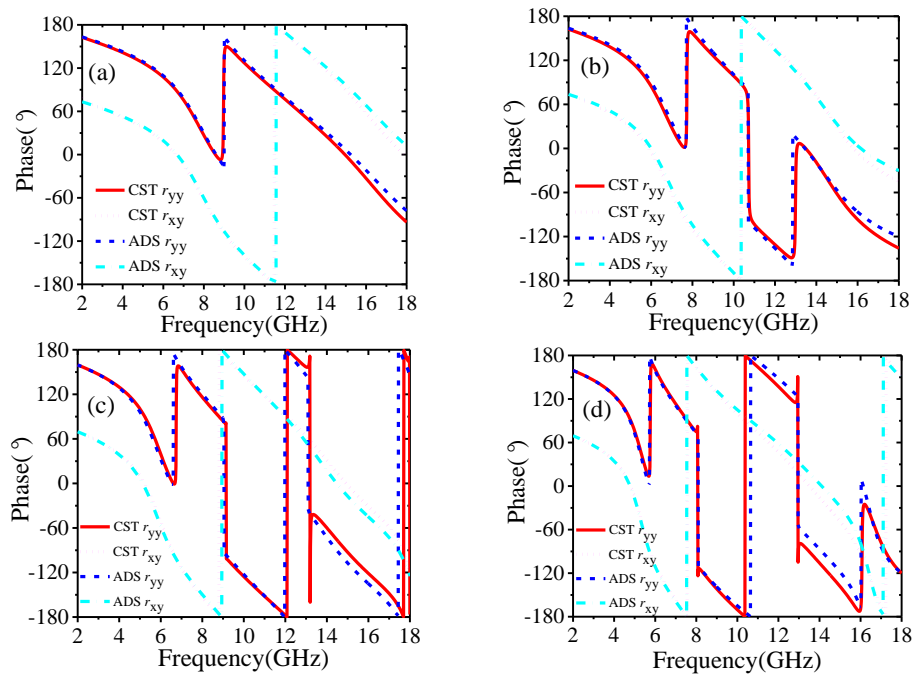

**Figure 5.** Circuital (ADS) and electromagnetic simulated (CST) cross-polarized reflected phase responses (a) unit cell 000; (b) unit cell 001; (c) unit cell 010; (d) unit cell 011.

## Application and simulation results in low-scatter patch antenna

### Antenna Structure

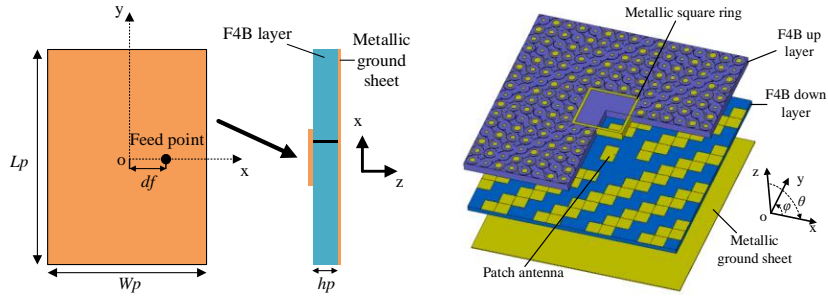

**Figure 6.** (a) top view and side view of the patch antenna; (b) the configuration of patch antenna combined with phase gradient metasurface consisting of 3-bit coding polarization conversion elements.

As shown in Fig.6 (a), a coaxial feed patch antenna is designed and its geometry parameters are as follows: $L_p$=12.4 mm, $W_p$=8.1 mm, $d_f$=3.0 mm, $h_p$=0.5 mm. According to the paper[15], the coding elements are arranged in the 2-dimensional phase gradient sequence to achieve the best monostatic RCS reduction. So we combine the patch antenna and the phase gradient metasurface (128×128 mm) consisting of 3-bit coding linear polarization conversion elements, to form a low RCS antenna, which is noted as antenna_I, as shown in Fig.6 (b). The patch antenna and metallic ground sheets of unit cell 000 and 001 are in the same horizontal plane. Thus, the antenna and metasurface share the F4B down layer with a thickness of 0.5 mm and the metallic groundsheet. The 4×4 unit cells above and around the patch antenna are removed and the metallic square ring (4×4 unit with 1.6 mm thickness and 4 mm height), is set around the patch antenna to protect antenna performance from the disturbance of metasurface. For reference, the patch antenna without metasurface is denoted as antenna_II.

## Radiation Characteristics

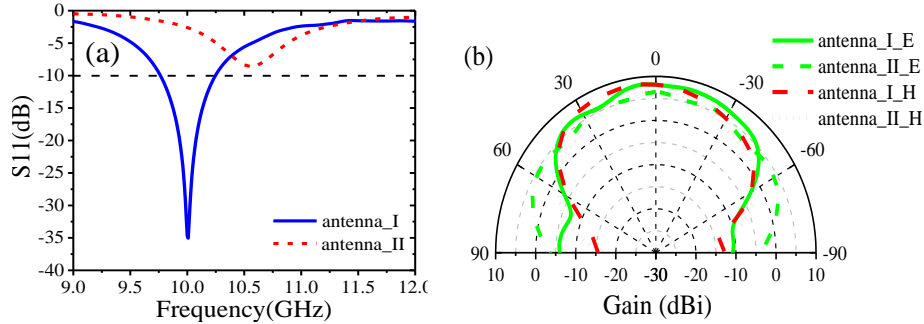

**Figure 7**. (a) The simulated reflection coefficients, (b) radiation patterns in the E-plane(xoz plane) and H-plane(yoz plane) of antenna_II at 10.0 GHz and antenna1 at 10.5 GHz.

The simulated reflection coefficients and radiation patterns of both antennas are shown in Fig. 7(a) and (b), respectively. The Fig. 7(a) reveals that the antenna_I operates around 10.0 GHz with -10 dB impedance bandwidth of 480MHz. In comparison, antenna_II works at around 10.5 GHz with poor impedance matching because the optimization of geometry parameters is done for the antenna_I without considering antenna_II. It can be seen that the beam width of antenna_I is narrower than antenna_II in both E- and H- planes. The gains of antenna_I at 10GHz and antenna_II at 10.5 GHz are 8.20 dBi and 6.47 dBi, respectively. Compared with antenna_II, the gain enhancement of antenna_I is attributed to the better impedance matching and increased antenna aperture due to the metasurface[8].

## Scattering Characteristics

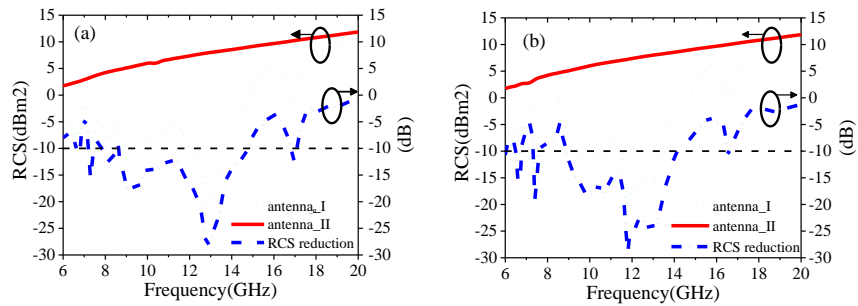

**Figure 8.** Simulated monostatic RCS of both antennas and RCS reduction of antenna_I compared with antenna _II versus frequency under normal incidence of (a) x-polarized wave, (b) y-polarized wave.

Fig. 8(a) and (b) shows the simulated monostatic RCS of both antennas for a normally impinging plane wave with x- and y-polarizations. Based on the simulated RCS results, we calculate the RCS reduction of antenna_I compared to the reference antenna, antenna_II, as also shown in Fig. 8. It is evident that the average 10 dB RCS reduction bandwidth is from 8 to 14 GHz and the RCS reduction value at around 13 GHz is at least 20 dB for both polarizations. Actually, the energy is redirected to the other orientation and converted to the orthogonal polarization by the phase gradient metasurface composed of linear polarization conversion elements[5]. Moreover, the antenna_I can realize RCS reduction with different levels from 6 to 18 GHz. The 3-bit coded unit cells have different amplitude and phase responses out of the operating bandwidth (8~14 GHz), which make contributions to the diffuse of the electromagnetic wave and further expand the RCS reduction bandwidth.

## Experimental verification

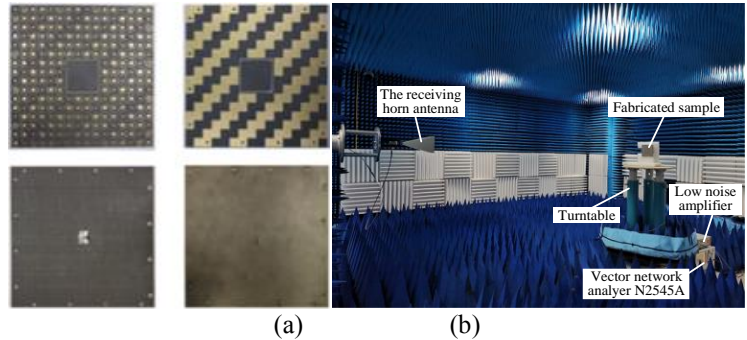

(a) (b)

**Figure 9.** (a) The top and bottom sides of two substrate layers of fabricated sample. (b) The experiment set up for the radiation pattern measurement of the fabricated patch antenna.

To validate the proposed design, the metasurface-patch antenna prototype is fabricated and its radiation performance is tested in a microwave anechoic chamber, as shown in Figs. 9(a) and (b). Actually, the down layer of proposed antenna is too thin to retain its original shape in the experiment, so we adjust the parameters of the patch in simulation as follows: Lp=12.2 mm, Wp=7.8 mm, df=2.8 mm. Besides, due to the thickness of soldering tin around the feed point on the patch in the experiment, the area between the upper and lower layers of F4B was filled with an air layer of 0.15 mm in the modeling of the CST simulation. Both measured and re-simulated reflection coefficients reveal that antenna_I operates at 10.4 GHz with -10 dB impedance bandwidth extending from 10.24 to 10.68 GHz while antenna_II at 10.9 GHz with a bad impedance matching in Fig.10(a). The measured gain of antenna_I at 10.4 GHz is 7.07 dBi, while of antenna_II at 10.9 GHz is 6.36 dBi, as shown in Fig. 10(b). Fig. 11 shows the normalized radiation patterns of both antennas. The measured results are consistent with those provided by the re-simulated results. We also test the scattering property of the fabricated prototype, as shown in Fig. 12(a). Limited by the testing condition, the bistatic RCS cannot be measured at present and only monostatic RCS reduction can be obtained, which is shown in Fig. 12(b). The slight change of the designed antenna structure has little impact on the monostatic RCS reduction. Both experiment and re-simulated results demonstrate that the proposed antenna could achieve a wideband RCS reduction from 6~18GHz and an average 10 dB RCS reduction from 8~14 GHz.

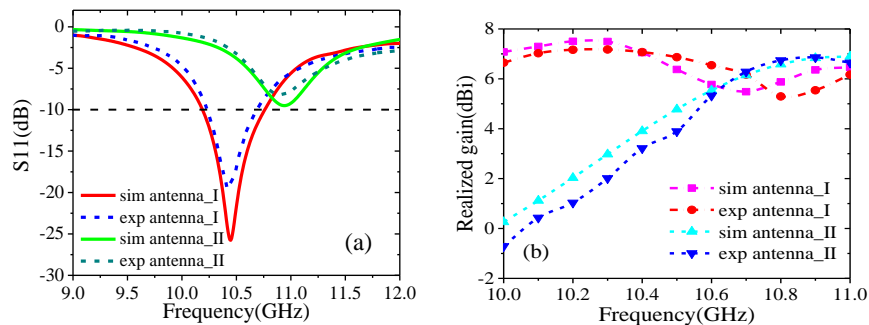

**Figure 10**. (a) The measured and re-simulated reflection coefficients, and (b) gain verus frequency of the proposed antenna(antenna_I) and reference antenna(antenna_II).

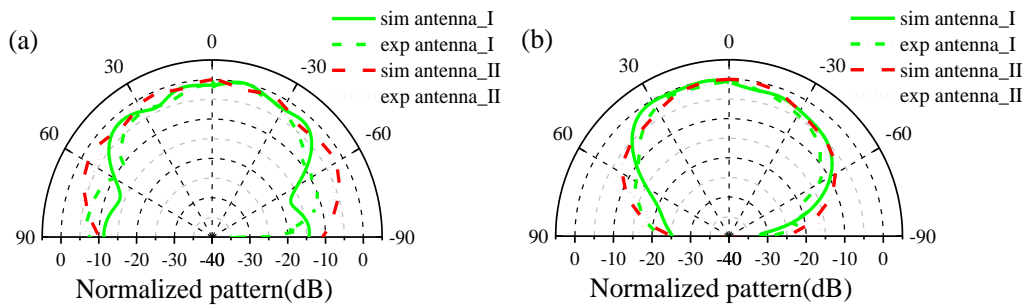

**Figure 11.** The measured and re-simulated normalized radiation patterns (a) in the E-plane(xoz plane) and (b) H-plane(yoz plane) of antenna_I at 10.4 GHz and antenna_II at 10.9 GHz.

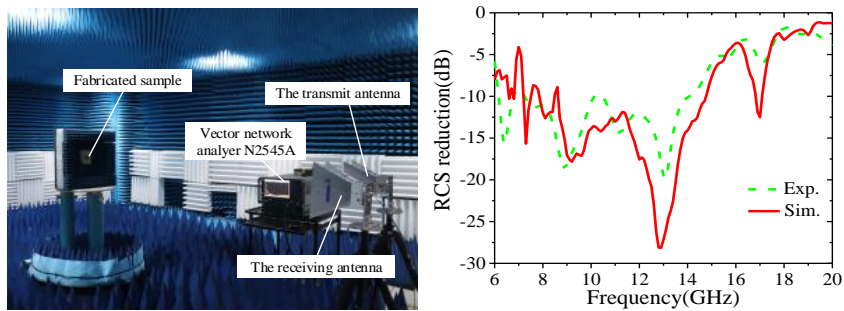

**Figure 12.** (a) The experiment set up for the monostatic RCS measurement of the fabricated patch antenna. (b) The measured and re-simulated monostatic RCS reduction of the proposed antenna compaerd to the reference antenna verus frequency under normal incidence of x-polarized electromagnetic wave.

## Conclusion

In this letter, a metasurface composed of 3-bit coding linear polarization conversion elements is designed. And its application to RCS reduction of the patch antenna is intensively studied. The 3-bit coding metamaterials are constructed by eight unit cells. Related ECMs of these unit cells are created to describe their electrical behavior for the two linear incident polarizations at the same time. A patch antenna is integrated on the metasurface composed of the coded elements placed with a 2-dimensional linear coding sequence. Both the simulation and experiment results demonstrate that the designed metasurface can largely reduce the antenna RCS at broadband from 6 to 18 GHz. In contrast, the antenna performances are not degraded significantly.

## Acknowledgments
This work was supported by the Fundamental Research Funds for the Central Universities (No. kfjj20190406), National Natural Science Foundation of China (61901217), Open Research Program in China's State Key Laboratory of Millimeter Waves (Grant No. K202027) and by the Postgraduate Research & Practice Innovation Program of Jiangsu Province under Grant (SJCX20_0070).

## Author contributions statement

X.K. and Q.W. conceived the original idea. Q.W. did the theoretical analysis and designed the metasurface. J.Y. and X.Y. fabricated the prototype metasurface. L.K. and S.J. performed the experiment. X.K. and X.Z. supervised the project. Q.W. and X.K. had deep discussion about this work and produced the manuscript.

## Additional information

**Correspondence** and requests for materials should be addressed to X.K.
**Reprints and permissions information** is available at www.nature.com/reprints.
**Publisher's note** Springer Nature remains neutral with regard to jurisdictional claims in published maps and institutional affliations.